\documentclass[a4paper]{article}
\usepackage{xcolor}
\usepackage{INTERSPEECH2021}
\usepackage{todonotes}

\usepackage{algorithmic}
\usepackage{amsthm}
\newtheoremstyle{colon}%
{}
{}
{\itshape}
{}
{\bfseries}
{:}
{ }
{}
\theoremstyle{colon}
\newtheorem{remark}{Challenge}
\newtheorem*{answer}{Focus Areas}
\usepackage{url}

\title{New Challenges for Content Privacy in Speech and Audio}
\name{Jennifer Williams $^1$, Karla Pizzi $^{2,3}$, Shuvayanti Das $^4$, Paul-Gauthier No\'e $^5$ }
\address{
  $^1$Electronics and Computer Science, University of Southampton, UK\\
  $^2$Fraunhofer AISEC, Garching, Germany;
  $^3$Technical University Munich, Germany \\
  $^4$Aflorithmic Labs Ltd., United Kingdom \\
  $^5$Laboratoire Informatique d’Avignon (LIA), Avignon Universit\'e, France}

\email{j.williams@soton.ac.uk, karla.pizzi@aisec.fraunhofer.de, shuvayanti@gmail.com, paul-gauthier.noe@univ-avignon.fr}

\begin{document}

\maketitle
\begin{abstract}
Privacy in speech and audio has many facets. A particularly under-developed area of privacy in this domain involves consideration for information related to \textit{content} and \textit{context}. 
Speech content can include words and their meaning or even stylistic markers, pathological speech, intonation patterns, or emotion. 
More generally, audio captured in-the-wild may contain background speech or reveal contextual information such as markers of location, room characteristics, paralinguistic sounds, or other audible events. Audio recording devices and speech technologies are becoming increasingly commonplace in everyday life. 
At the same time, commercialised speech and audio technologies do not provide consumers with a range of privacy choices. 
Even where privacy is regulated or protected by law, technical solutions to privacy assurance and enforcement fall short. 
This position paper introduces three important and timely research challenges for content privacy in speech and audio. 
We highlight current gaps and opportunities, and identify focus areas, that could have significant implications for developing ethical and safer speech technologies.
\end{abstract}
\noindent\textbf{Index Terms}: speech privacy, content masking, privacy evaluation, speech recognition, speaker recognition

\section{Introduction}
The idea of privacy in audio data is not clear-cut even though privacy, along with security, are both at the core of individual protections in the digital age \cite{nautsch2019gdpr}. Privacy and security are often conflated, especially in the legal domain \cite{bambauer2013privacy}. In this paper, we adopt the stance that \textbf{privacy} is related to \textit{controlling access to information} whereas \textbf{security} involves \textit{how such information could be used (or misused)}. It is an open question of how to ensure data privacy for any dataset, especially for speech and audio data. For audio data, there are many different kinds of private information contained in a recording. Background noise may indicate a geographical location, voices may be captured from people who have not consented to sharing their speech data, and a recording could even contain person-related information. In this paper, we are especially interested in speech content privacy, though we recognise that speech privacy is intertwined with the more general case of audio processing.  

Early efforts to address speech privacy involved a technique of using meaningless masking noise to obscure speech information, effectively making speech unintelligible to human eavesdroppers \cite{tamesue2006study}. This foundational idea of noise masking often assumes that human talkers are inside of a room (and the room characteristics are known), and that it is desireable to conceal entire conversations. The idea of noise masking persists as more recently it is used to provide privacy for audio captured by smartphones \cite{tung2019exploiting}. Noise masking is conceptually related to privacy achieved through microphone jamming \cite{chen2022big}. However, both of these techniques (noise masking and jamming) make assumptions about the audio capture conditions as well as an intended eavesdropper (more accurately referred to as an \textit{attack vector}). The techniques also seek to provide a type of blanket privacy, and do not allow fine-grained control or the opportunity for interlocutors to choose privacy levels. 

\begin{figure}[h!]
    \centering
    \includegraphics[width=0.45\textwidth]{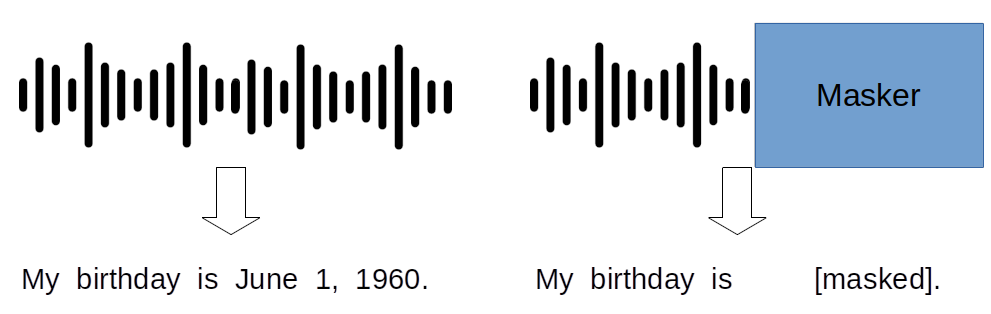}
    \caption{Concept diagram of applying a technical masking solution to hide or conceal content in audio. 
    On the left: speech recognition without a masker; on the right: speech recognition with masker.
    The portion of audio containing birth date information is masked in such a way that an automatic speech recognition system could not transcribe it.}
    \label{fig:schematic_masker}
\end{figure}



\begin{table*}[h!]
    \centering
    \caption{Different types of information interception in human and machine interaction.
    Here, data is transmitted between partners (e.g., during a conversation).
    This data is intercepted through an \emph{information breach}.}    \begin{tabular}{cllp{7cm}}
    \hline
        \textbf{No} & \textbf{Data Transmission} & \textbf{Privacy Breach} & \textbf{Example} \\
         \hline
         1 & Person to Person & Person & In any public space, one can overhear lots of conversations (e.g., on a bus or in a restaurant). \\
         2 & Person to Person & Machine & Having a conversation in close proximity to a smart device can potentially allow the device to listen, if certain activation words are mentioned or speech is (intentionally) recorded in the background. \\
         3 & Person to Person (via Machine) & Machine & When using a real-time translation program, the device could eavesdrop or intercept private information from a human-to-human conversation. \\
         4 & Person to Machine & Person & With the spread of smart assistants, more and more actions are performed based on voice orders.
         Eavesdropping them can potentially also reveal private information (e.g., a family shopping list or the names of people to call and their phone numbers). \\
         5 & Person to Machine & Machine & \emph{Case 1:} Wake-word detection that is used for activating devices often do not differentiate between commands addressed directly to the device from other natural conversation. These other conversations could be human-to-human but also human-to-machine. \\
         & & & \emph{Case 2:} If audio data containing private information is part of a training dataset, then a machine learning algorithm might be able to incorporate this private information into a model (which could possibly be extracted from the model later on).\\
         6 & Machine to Person & Person & When using a voice-based assistant, often some parts of the information are synthetically spoken out loud (e.g., when using a navigation program or using a screen reader assistive device). \\
         \hline
    \end{tabular}
    \label{tab:eavesdropping}
    \vspace{1mm}
\end{table*}

In this paper, we are concerned with privacy for content in speech and audio. We use the term \textit{content} to distinguish content/context information from the narrower definition of speaker voice characteristics. In doing so, we aim to distinguish content privacy from voice privacy as in the VoicePrivacy Initiative \cite{tomashenko2020introducing}. The VoicePrivacy Initiative was established in 2020 and at this time it currently is focused on anonymisation of speaker identity. Instead, we are concerned with information in speech and audio that is in the same spirit of semantics. For example, if a person uses words to reveal their name, birthday, address, or bank account number, then rendering their acoustic voice characteristics anonymous would not be sufficient to conceal this type of private information. Consider the conceptual schematic in Figure~\ref{fig:schematic_masker}. In this example, a masker is applied to speech to conceal a specific word (a birth date) in an utterance. We will discuss technical masking solutions in more detail in Section~\ref{sec:masker_types}. More generally, content information may stretch to include paralinguistic audio events as in styles of laughter or disordered speech patterns like stuttering. 

From a consumer perspective, speech technology has become increasingly commonplace at the time of this writing. Governing bodies such as the EU \cite{voigt2017eu} and UK \cite{UKparliament} have passed legislation or are conducting additional investigation on the need for privacy protections for citizens. Further, there are known cases of unauthorised recording from audio-enabled devices where data is provided to unauthorised persons, raising concerns of how far privacy protections must reach, both legally and technically \cite{griffin}. Speech data that is captured by devices has also been sought out as part of enquiries into serious violent crimes \cite{cuthbertson}. If such recordings were admissible as evidence in a court, it may be desireable to admit partial portions of a conversation. The seizure of data from devices also has implications for data governance, as the data (or copies of it) would then be stored at additional locations. This may affect enforcement of a person's right to be forgotten \cite{rosen2011right}.


The threat landscape for content privacy issues in speech and audio is extremely broad in terms of how data is transmitted and how information can be breached. Each case can be considered unique, therefore it is challenging to arrive at a unified viewpoint of the requirements for speech privacy. In Table~\ref{tab:eavesdropping}, we provide an overview of different situations related to speech privacy. In some scenarios the risk for harm is elevated (e.g., case No. 2 and 3 where speakers have an inherent expectation of privacy), even when an information breach requires sophisticated knowledge or technical skills to carry through. 

If work toward content privacy does not begin to mature within the speech and audio research community, it will become very difficult to enforce the legal protections that are afforded to individuals or to protect confidential business information. This is due in part to the sophistication of technologies that are able to violate privacy \cite{shi2022speech}, in addition to unintentional privacy violations \cite{kwong2019hard}. The harms are especially amplified for vulnerable groups and those who rely on speech technology as an assistive device. In the end, everyone benefits from content privacy. 

This paper delivers three directions for content privacy research and explains the challenges as well as some proposed focus areas. The three directions are: (1) technical solutions for content privacy, (2) impacts on downstream tasks, and (3) standardising content privacy evaluation. We encourage the speech and audio community to incorporate aspects of these research directions into their current and future work. Such an important topic should not be left to industry-driven interests or stymied by missing innovation. 


\vspace{-2mm}
\section{Background}
Some of the earliest work on speech privacy addressed the scenario of preventing a human eavesdropper from intercepting information from overhearing a conversation. In such cases, the approach to conceal conversational content was focused on ensuring that all audible speech was made unintelligible. For example, in \cite{tamesue2006study} a masking noise was generated and evaluated along with a sound pressure level to indicate privacy levels. In the work of \cite{chen2008audio}, a speech re-synthesis technique was used to apply a vocal tract transform function and to replace speech regions with re-recorded vowels. These techniques for achieving content privacy could be developed further to address a limited number of variations of our privacy scenario No. 1 in Table~\ref{tab:eavesdropping}. For large collections of speech transcripts, language models have been shown to be useful for detecting classes of sensitive words for hate speech \cite{mozafari2019bert}, and this same idea could be developed for detecting other classes of sensitive private information that needs to be masked. In fact, the BERT model \cite{kenton2019bert} already learns from masked data. The notion of learning from masked data could potentially be expanded to create new types of privacy-preserving language models.

Blanket approaches to content privacy are often evaluated with metrics that measure intelligibility from a human perspective \cite{tamesue2006study}, and more recently by reporting automatic speech recognition (ASR) word error rates \cite{chen2008audio, dong2022adversarial}. But evaluating privacy depends on the mode of data transmission and type of privacy breach. Existing standards for privacy measured through intelligibility are discussed in more detail in Section~\ref{sec:standards}. Work from \cite{cychosz2020longform} proposes guidelines for ethical audio data collection from a behavioral sciences perspective. They call for the research community to explore more technical solutions to privacy, for example developing techniques to extract features from audio before storing it, in order to avoid having to store recordings. 

Privacy in audio may involve developing algorithms that conceal private information while also retaining other useful information for tasks like classification of environmental sounds \cite{chen2008audio, kumar2015sound}. Similar work has been developed for cough detection \cite{larson2011accurate} where audio is recorded but any speech that is incidentally captured can be rendered unintelligible. More recently, \cite{liang2020characterizing} surveyed people's perception of privacy before and after audio degradation. The purpose of degradation was to make speech information unintelligible while preserving enough information in the audio to perform activity recognition for common household tasks (washing dishes, cooking food, bathing/showering, etc). They found that intentional degradation of audio significantly improved people's perception of privacy, while still performing well on the classification tasks. However, the study did not assess whether speech was still recognizable though an ASR system or how people's perception of privacy would change if they had seen an ASR transcription of degraded audio. 

Non-speech classification tasks continue to become more prevalent even though not all researchers are actively addressing privacy issues related to audio capture. An example is ``BodyScope'' from \cite{rahman2014bodybeat} who developed a technique for detecting activities from wearable audio sensors (e.g., microphones) and used the captured audio for classifying sound into events such as eating, drinking, speaking, and laughing. Striking a balance between masking private information (such as speech) from audio capture in general is very difficult. Even recent advances in microphone jamming do not address fine-grained control of masking \cite{chen2022big}. While jamming may conceal human speech, it can sometimes fully degrade audio that would have been useful for other tasks.



\section{Challenges}
In this section, we present three of the most important challenges to achieving forward progress for content privacy in speech and audio. Addressing all of these challenges will be necessary for extending the current state of the art and addressing technical gaps that persist, including attitudes about the importance of privacy. For each, we identify a challenge question followed by proposed focus areas and discussion.

\subsection{Technical Solutions for Content Privacy}
\label{sec:masker_types}
With the first challenge, we refer to the situations No. 2 -- 6 from Table~\ref{tab:eavesdropping} very generally.
We ask for more precise definitions to define the attack scenarios in order to formulate the technical requirements for a masking algorithm.
\begin{remark}\textbf{What types of algorithms will facilitate content masking in speech and audio, and how does this change depending on the type of content information to be masked?}
\end{remark}
\begin{answer}\textbf{
First, we propose to establish a taxonomy for content privacy in order to define suitable protections for attack scenarios. 
With this at hand, in the best case, we could develop new masking algorithms that potentially generalise across multiple types of content privacy scenarios (e.g., masking sensitive fine-grained content like spoken birth dates as well as short bursts of unintentionally captured third-party conversation in the background).
}
\end{answer}

In the following, we use the term \emph{masking} to refer to any type of method used to obscure information (e.g., applying a special noise or modifying signal-level features).
The mask could be audible (to conceal speech for human listeners) or subsonic (for machine listening), and it can obscure speech content as well as non-speech content that contains private information.  
The main idea is to make the content unintelligible to human and machine eavesdroppers.
In order to apply masking algorithms to speech data, we see various points that should be covered in a suitable taxonomy:
\begin{itemize}
    \item How is privacy different for speech versus non-speech audio? $\rightarrow$ Human speech may contain very different types of private information, including health information or personal content.
    Hence, the requirements for speech might be even more strict than for general audio data. Sensitive audio information may contain geographical identifiers that are either localised in the speech signal (a car horn) or global (room characteristics). 
    \item Are there limits to what can be masked?
    $\rightarrow$ It seems impossible to mask all private content at the same time. For example, if we try to mask private speech content, background noises and emotions at the same time, we might loose too much data. Currently there is no way to quantify or characterise such data loss. 
    \item What is the granularity of control that we want to have and why? How do the granularity needs affect design decisions? $\rightarrow$ Masking out complete words seems a very intuitive solution. 
    However, sometimes already parts of a word might be enough to ensure an appropriate level of privacy (e.g., for someone born in the 20th century, masking just the decade and the precise year).
    \item 
    Are there some recent machine learning (ML) trends that we can take advantage of? $\rightarrow$ In ML, privacy has evolved as a major topic, covering different attack scenarios like model inversion \cite{fredrikson2015model} and membership inference \cite{shokri2017membership}.
\end{itemize}

\noindent These questions may be considered from different perspectives, including a software approach (algorithm-based privacy) or hardware or edge solutions (e.g., embedded in the audio capture device). And in some cases, the solution will require a combination of both software and hardware, such as computing at the edge (i.e., disconnected from the cloud) or through digital signal processing (DSP) techniques on-device. 
More concretely: let us consider content masking in automatic speech recognition (ASR). 
If the data includes someone mentioning their birthday, for some applications it could be enough to add additional noise on the portion of audio that reveals a date.
If the ASR model then recognizes a different birthday because the underlying language model predicts that a date occurs in the sequence of words, that could still be considered acceptable (because the original private information was concealed).

In other circumstances, we might want the model to not transcribe any false data and rather leave out specific words. In this case, one could benefit from complete silence masking. Hence, one needs to consider different types of masking depending on the requirements. If a reversible masking is desireable, then the technical masking solution would also require a ``key'' to undo a mask after it has been applied. Reversible masking may have uses in real-time speech transmission or scenarios where a very sensitive speech database must be shared with an authorized party.

Clearly there is not a one-size-fits-all masking solution for audio data. 
Which masker to choose depends on whether we are masking some speech in a database, speech in real-time (like telephony or video-telephony), or speech while compressed, etc. And it may also depend on the type of information being masked such as words in speech, audio event noises, or background speakers. Further, we need to consider whether or not the masked audio/speech needs to be recoverable/reversible.

\subsection{Impacts on Downstream Tasks}
The second challenge mostly refers to situation No. 5 from Table~\ref{tab:eavesdropping}, but could potentially be generalized to situations No. 2 and 3 as well.
We mainly address the issue of models building on top of other models (or models built upon data that has been significantly transformed).
If a privacy solution modifies audio through masking, we need to consider the implications for developing subsequent models for other tasks that are based on this pre-processed data.
\begin{remark}\textbf{After concealing different types of informational content in speech or general audio, how can we anticipate and prepare for the impact on downstream applications such as database sharing, speech recognition, speaker verification, or audio scene analysis?}
\end{remark}
\begin{answer}\textbf{In order to overcome problems in downstream tasks, different models of a pipeline should not be considered in terms of modularity (i.e., in isolation).
Hence, we propose to create shared tasks or challenge tasks that provide a suite of baseline tools corresponding to example downstream tasks that can be used for benchmarking. 
}
\end{answer}

The problem in this domain involves finding an optimal privacy policy while preserving just enough information for the audio to be useful in a task. 
The most commonly used downstream tasks in all previous work on this topic are ASR and automatic speaker verification (ASV). 
Beyond this, there are plenty of other downstream tasks that use audio data as input, for example for health analysis or activity recognition. 
With model pipelines, new questions arise:
\begin{itemize}
    \item If some of the audio is masked (e.g., words or other content features), how does that affect certain tasks? $\rightarrow$ For example, if we remove markers of room or location information can we still conduct audio scene analysis? 
    Or if we remove background speech from children on a telephony or video-telephony call, does this make it more challenging to perform other tasks such as speech compression or speech-to-speech translation? 
    One main challenge is that we do not always know what the masked speech/audio will be used for later on. 
    \item What are some strategies for estimating the impact of a technical masking solution, including what can be done later with the audio at other stages in the pipeline? $\rightarrow$ We believe that a masker should always be chosen with at least one downstream task in mind. However, it would be useful for future research to determine if there is a more universal approach to first prioritising privacy (i.e., the content to be masked) and then applying the downstream tasks without any loss in performance, after sensitive content has been concealed.
    \item Is there a way to create content privacy solutions that are reliable enough for very sensitive speech (e.g., speech used in medical research) to be useful in other tasks or at least shared between institutions without having to be stored on special servers while protecting the rights of individuals? $\rightarrow$ This is particularly important for large databases of speech, especially ones that might be considered sensitive (e.g., medical child speech \cite{das2022speech}). Increased data sharing, when done properly, can result in large gains for people who stand to benefit most from speech and audio technology. 
\end{itemize}

\noindent For some of the above questions, a potential solution is to incorporate privacy-preserving mechanisms directly into a downstream task or model, rather than relying solely on data transformation. An example of this could be developing an ASR system that has been trained on data where protected content has been masked, possibly by adding an acoustic marker in the signal to indicate that the ASR system must skip or ignore sensitive speech content. Such a system may perform better than one that has been trained on unmasked speech. For example, an ASR system could be trained to learn that credit card information should never be recognised or transcribed. 
On the other hand, to stay with our example, if one wanted to conduct speech-based online shopping then credit card details are necessary. We argue that this should be a choice for those who use speech technologies. Further, an ASR system that has privacy-by-design would need to be adopted by industry in order to be included in commercial products.





\subsection{Standardising Content Privacy Evaluation}
\label{sec:standards}
The third challenge very generally refers to notions of privacy in audio data, hence, all situations from Table~\ref{tab:eavesdropping} should be considered here.
\begin{remark}\textbf{How can we develop efficient and objective measures of content privacy that can be used for privacy assurance and compliance?}
\end{remark}
\begin{answer}\textbf{We propose that the speech and audio research community establishes a new international working group that can collaborate on developing a new set of international standards and measurements that address aspects of fine-grained content privacy that are currently missing from existing standards. }
\end{answer}

Here we expand on some of the considerations for developing standards. These ideas can be developed further as part of an independent research agenda designed to address speech content privacy, or they could be added to existing and ongoing research efforts (e.g., dataset creation, training new ASR systems etc). 

\begin{itemize}
    \item What are the desireable characteristics of a content privacy metric? $\rightarrow$ Other speech standards are easy to use for assessing speech. The algorithms run quickly, results are reliable, and the metric is often interpretable. From the first challenge in this paper, a taxonomy of content privacy would be helpful as different types of content may require different metrics. Further, it may be useful to have metrics that assess algorithms and ML systems that can assure privacy, beyond data transformation.
    \item What are some special cases where audio privacy is particularly difficult to assess with a standard metric? $\rightarrow$ Some types of content may be considered sensitive, but it would be difficult to establish a ground truth for comparing if information has or has not been effectively removed. For example, there is high variation for paralinguistic events (e.g., laughing, crying, screaming, etc) though this information could easily be used to identify a person in audio. It may be challenging to establish standard metrics for assessing masking of speech and audio events that have naturally high variation. 
    \item What are the similarities and differences for standard privacy metrics for content privacy versus speaker anonymisation? $\rightarrow$ In a different form of privacy, focused on concealing acoustic voice characteristics of speakers, one goal is to measure how close or similar two speaker voices are in terms of acoustics. It may be possible to build on this idea for content privacy, perhaps through developing a new type of audio embedding that can be used for comparisons before and after masking. One challenge with this approach is that content privacy aims for fine-grained masking (in the time domain or frequency domain) whereas audio embeddings are usually learned from a large temporal context. 
\end{itemize}

\noindent Numerous privacy metrics already exist for the speech community. A list of these privacy metrics is provided in \cite{stout2015speech}. While the existing metrics have proven to be very useful, they are often based only on perceived intelligibility (e.g., as a human eavesdropper, our scenario No. 1, 4, and 6 in Table~\ref{tab:eavesdropping}), and measured at a very global level for entire conversations. They do not address very fine-grained privacy such as concealing specific words or phrases or sensitive non-speech events. In \cite{bohn2000navigating}, it is argued that anywhere there is audio there are standards, therefore we have standards everywhere. The particular standards introduced in \cite{bohn2000navigating} are for audio-related hardware and equipment and not directly relevant to this paper, but the point is taken that many standards already exist. 

Privacy index (PI) is used for open office acoustic environments \cite{muller2019relationship}. More suitable to closed rooms is the STI (speech transmission index) \cite{payton1999method} which holds across many conditions such as noise, reverberation, and speaking style. Speech privacy class (SPC) \cite{bradley2010new} is another standard privacy metric. The speech intelligibility index (SII) is an ANSI standard (e.g., American National Standards Institute) \cite{pavlovic2018sii}. The work of \cite{larm2006experimental} performed a comparison between rapid STI (RASTI), STI, and SII and reported that STI and SII are comparable for measuring intelligibility. In all of these metrics, it is not possible to apply them in fine-grained scenarios, and further they are oriented toward intelligibility and do not apply for paralinguistic content privacy.

Recent efforts to address the need for a new content privacy metric come from \cite{williams2021revisiting} proposing a masking error rate (MER). For that metric, they assume that content privacy refers solely to words and that a ground truth transcription of speech (i.e., unmasked content) is available. Given these assumptions, the metric calculates proportions of words that are correctly or incorrectly masked and produces a single value score. It is not clear how this metric could be applied in a real-world scenario, as the work only proposed the metric but did not use it in a task. Furthermore, the MER only applies to words and does not address how it can be adapted (if at all) for paralinguistic content privacy or non-speech events. 




\section{Discussion}
We have presented three challenges for the speech and audio research community to begin exploring in order to advance a critically under-explored area of privacy. It is important that privacy levels are treated as a choice for individuals to decide. Toward this goal, we have discussed some of the requirements for developing technical solutions. Following that, we discussed how content privacy could be considered in terms of downstream tasks. Finally, we have asked for the speech and audio community to come together and begin developing international standards for content privacy. We have provided a set of issues that can be considered toward such standards. 

It is known that tech companies store voice information from consumers who use their products \cite{whitaker, field}. These companies have also acknowledged sharing data with third parties, causing harm to consumer privacy. Hence, consumers are often left with limited options: they can use a conveniently good and available product risking their privacy \cite{9546911} or not use such products and forego the benefits of speech technology. 

Even with security and privacy protection mechanisms in place, we often find adversarial development cycles in the research and development domain. Once a protection technique has been created and published, hackers (or other academics) try to point out the pitfalls. In order to avoid putting privacy at risk in the first place, products should not be commercialised unless a company has evaluated the privacy aspects and released this information to consumers before making products available for purchase. Ideally, academics and interested people will then still probe commercial products for undetected privacy weaknesses and further make that information publicly available. What exactly needs to be told to consumers is not clear. As this comes at high costs for the companies, legislators are beginning to pay more attention to this issue \cite{zaeem2020effect,vanberg2021informational}. Companies are rarely motivated to incorporate privacy features in their tech unless there is a law requiring this.

We propose that researchers adopt a stance towards speech privacy and security that is analogous to recent trends of addressing issues of algorithmic bias. We encourage researchers to include statements of how privacy and security have been considered in their work. We further expect an increased discussion of regulation and policy to hold technology developers accountable for offering more privacy options in commercial products.

\section{Acknowledgements}
This research was supported by the Bavarian Ministry of Economic Affairs, Regional Development and Energy.

\bibliographystyle{IEEEtran}
\bibliography{mybib}
\end{document}